\documentclass[aps,prl,amsmath,amssymb,floats,preprint,showpacs]{revtex4}

\usepackage[dvips]{epsfig}

\begin{document}

 \title{Nonequilibrium quantum decay and decoherence in quantum
impurity problems}
 \author{Holger Baur$^1$, Andrea Fubini$^2$, and Ulrich Weiss$^1$}
 \affiliation{$^1$Institut f\"ur Theoretische Physik, Universit\"at Stuttgart,
   D-70550 Stuttgart, Germany}
 \affiliation{$^2$Dipartimento di Fisica, Universit\`a di Firenze,
    and INFM, UdR Firenze,  Via G.Sansone~1, I-50019 Firenze, Italy}
 \date{\today}

 \begin{abstract}
Using detailed balance and scaling properties of integrals that appear in 
the Coulomb gas reformulation of quantum impurity problems, we establish 
exact relations between the nonequilibrium quantum decay rates of the 
boundary sine-Gordon and the anisotropic Kondo model at zero temperature.
Combining these results with findings from the thermodynamic Bethe
ansatz, we derive exact closed form expressions for the quantum decay rate
of the dissipative two-state system in the scaling limit.
These expressions illustrate how the crossover from weak to strong tunneling
takes place. We trace out the regimes in which the usually applied Golden Rule 
(nonadiabatic) rate expression fails. Using a 
conjectured correspondence between the relaxation and dephasing rate,
we obtain the exact lower bound of the dephasing rate as a function of
bias and dissipation strength.  
 \end{abstract}

 \maketitle

Quantum impurity problems (QIPs) have attracted a great deal of interest 
recently. This is because the underlying physics is non-trivial and the models
are manageable technically despite their essentially nonperturbative nature.
In addition, they have a multitude of experimental applications, including
the Kondo effect, quantum dots, dissipative quantum mechanics, tunneling in
quantum wires and fractional quantum Hall devices \cite{weissbook}. 

There have been discovered various relations between 
thermodynamic quantities of the 
anisotropic Kondo model (AKM) and the boundary sine-Gordon (BSG) field theory
model \cite{fendley95a}. 
Each of these integrable models is of considerable interest and it is 
remarkable that they are closely related. 
The AKM corresponds to the dissipative 
two-state system (TSS) in the Ohmic scaling limit 
\cite{leggett87} and the BSG model corresponds to the Schmid 
model \cite{schmid83}. 
The latter describes a quantum Brownian particle coupled to an Ohmic 
heat bath and moving in a tilted cosine potential.
The tight-binding (TB) limit of the Schmid model is equivalent to the 
strong-backscattering limit 
of the BSG model, while the weak-binding (WB) limit of the Schmid model
represents the weak-backscattering limit of the BSG model. 
In the sequel, we use the language of dissipative quantum mechanics.

The equivalence or difference of these
models can most easily be seen in the Anderson-Yuval Coulomb gas 
representation for the partition function. The partition function of
all these models can be expressed 
as the partition function of a one-dimensional classical gas of positive and 
negative unit charges with ''log-sine'' interactions. The AKM (or TSS) 
problem differs from the BSG (or Schmid) model by a different ordering 
prescription.

In this Letter, we calculate the full nonequilibrium quantum decay rate
of the TSS at zero temperature in closed analytic form. 
Upon using a conjectured correspondence
between relaxation and decoherence in the TSS, we also determine the
lower bound for decoherence in this system. 
These outstanding problems are solved
by establishing exact functional relations between nonequilibrium quantum 
rates of the TSS and the Schmid model.

The TSS Hamiltonian in pseudospin form reads 
\[
H_{\rm TSS}^{} = - {\textstyle \frac{1}{2}} \Delta_{\rm T}^{}\,\sigma_x^{} 
- {\textstyle \frac{1}{2}}\,[\,\epsilon + f(t)]\,\sigma_z^{}
 \; ,
\]
where $ \hbar = k_{\rm B}^{}=1$.
Here $\Delta_{\rm T}^{}$ is the tunneling coupling, $\epsilon$ is the bias 
energy, and $f(t)$ is a random force with Gaussian statistics. 
All effects of $f(t)$ are captured by the second integral of the
force autocorrelation function, denoted by $Q(t)$. In the scaling
(or field theory) limit, we have
$Q(t) = 2K \ln[\,(\omega_{\rm c}/\pi T)\sinh(\pi T |t|)\,] 
+ i\,\pi K\,{\rm sgn}(t)$, where $T$ is temperature and $\omega_{\rm c}$ is a 
cut-off frequency. The Kondo parameter $K$ is a dimensionless Ohmic damping 
strength, which is the inverse of the Luttinger parameter $g$ or the filling 
fraction $\nu$ in quantum impurity problems.

The TB Schmid model in second quantized form with coupling energy 
$\Delta_{\rm S}^{}$ and bias energy $\epsilon$ is given by
\[
H_{\rm S}^{} = - {\textstyle \frac{1}{2}}
\Delta_{\rm S}^{}\! \sum _n \left( a_n^\dagger 
a_{n+1}^{} + {\rm h.c.} \right) 
- [\,\epsilon + f(t)]\! \sum _n \! n a_n^\dagger a_n^{}   \; .
\]

The nonequilibrium dynamics of both models
may be computed using the Keldysh or Feynman-Vernon formalism
for the reduced density matrix (RDM).
Consider a particular path on the $(q,q')$-plane of the RDM parametrized 
by charges $\{ u_j=\pm 1\}$ and $\{v_i=\pm1\}$ chron\-ologically arranged 
on the $q$ and $q'$ path, respectively. The influence functional due to the 
stochastic force $f(t)$
for $k$ charges on the $q$-path and $\ell$ charges on the $q'$-path, divided up
into the self-interactions of the paths and the interaction between the paths,
reads
\begin{eqnarray*}
\cal{F} &=& \exp\bigg\{\sum_{j>i =1}^k u_j u_i Q(t_j - t_i)  \\
&+&    \sum_{j>i=1}^\ell v_j v_i Q^\ast (t_j' - t_i') 
- \sum_{j=1}^\ell \sum_{i=1}^k v_j u_i Q(t_j' -t_i) \bigg\}\, .
\end{eqnarray*}
For the TSS the charges $u_j, v_i$ 
alternate in sign, while for the Schmid model they may be ordered
arbitrarily.
At long times, the Schmid model is a Poissonian transport model: the
population dynamics results from uncorrelated direct forward  and backward 
transitions by $n$ wells, $n =1,\, 2, \cdots$.
Since they are not classical, the respective weights per unit time (``rates'')
$\gamma_n^\pm$ are not necessarily positive because of quantum interference 
\cite{saleurweiss01}.
In the TB representation,
the Fourier transform of the population distribution
$\hat{P} (k,t) = \sum_n e^{ikn}_{}\, P_n(t)$ takes the form
$\hat{P} (k,t) = 
\prod_{n=1}^\infty \exp\left[t\left(e^{ikn}_{} -1\right)\gamma_n^+
+ t\left(e^{-ikn}_{} -1\right)\gamma_n^- \right]\,$,
where the rates $\gamma_n^\pm$ describe uncorrelated direct 
forward/backward transitions by $n$ TB states. 

The Poissonian dynamics is found upon carrying out
a cluster decomposition of the individual path contributions to the
Laplace transform $\hat{P}(k,\lambda)$ in 
the limit $\lambda\to 0$. A cluster is a $\lambda$-independent 
(irreducible)  path section which starts and ends in a diagonal state of the 
RDM, $\sum_i  u_i = \sum_j v_j $ and cannot be factorized 
into clusters of lower order.  By definition, the clusters are 
noninteracting. Therefore a succession of clusters factorizes and the 
integral over a (time) 
interval separating any two clusters produces a factor $1/\lambda$.
Path segments which start and end in a diagonal state and again have
interim visits of diagonal states are reducible, i.e. they
factorize into clusters of lower order and factors of 
$1/\lambda$. However, after subtraction of all 
the reducible components, an irreducible part is left. The irreducible 
contributions of all charge arrangements with excess charge
$\sum_j u_j = \pm n$, $\sum_i v_i = \pm n$ define the transition rates 
$\gamma_n^\pm$.

The irreducible parts of arrangements with $2(n +\ell +k)$ charges
contribute to the order 
$\Delta^{2(n+\ell+k)}$ of $\gamma_n^+$. They can be divided into a
multitude of different subsets. 
Each subset $\alpha$ is a particular arrangement of $n+ 2\ell$ 
time-ordered $u$-charges and $n+2k$ time-ordered $v$-charges. 
The number of subsets with different order of the $u$- and $v$-charges is 
$N = N_u N_v = \frac{(n+ 2\ell)!}{\ell ! (n + \ell)!}
          \,       \frac{(n+ 2k)!}{k!(n + k)!}$.
In each individual subset $\alpha$, all possible orders of the $u$-charges 
relative to the $v$-charges are taken into account, which amounts to
$N_{u,v} = \frac{(2n + 2\ell + 2k)!}{(n+2\ell)!\,(n+ 2k)!}$ different
arrangements. Each subset $\alpha$ defines a partial forward rate 
$\gamma_n^{\alpha,+}$. The corresponding partial backward rate 
$\gamma_n^{\alpha,-}$ is represented by the arrangement 
with reverse order and reverse sign of the charges.
The individual partial rates are directly related by detailed 
balance, $\gamma_n^{\alpha,-} = e^{-n\,\epsilon/T }\,\gamma_n^{\alpha,+}$.
Detailed balance is satisfied for every partial rate
because the interaction $Q(z)$ between 
the $u$- and $v$-charges is analytic for complex time $z$ in the strip
$0 \ge {\rm Im}\,z > - 1/T$ and has the property 
$Q(t- i/T) = Q^\ast (t)$. 

In the remainder, we restrict the attention to  zero temperature. 
At $T=0$, all frequencies of the Schmid and TSS model can be absorbed 
into a single dimensionless parameter,
$x_{\rm S}^{} = \left(\epsilon/\omega_{\rm c}\right)^K
\,\Delta_{\rm S}/\epsilon$ and 
$x_{\rm T}^{} = \left(\epsilon/\omega_{\rm c}\right)^K
\,\Delta_{\rm T}/\epsilon$.
For later convenience we also introduce the frequency scale
$\epsilon_0^{}$ (analogous to the Kondo scale in the Kondo model
and the scale $T_{\rm B}'$ in QIPs \cite{fendley95b})
and the dimensionless bias $v = \epsilon/\epsilon_0^{}$. The scale 
$\epsilon_0^{}$ is related to the bare parameters of the Schmid and 
TSS model as \cite{weissbook}
\[
\epsilon_0^{2-2K} = \frac{2^{2-2K}_{} \pi^2}{\Gamma^2(K)}\,
\frac{\Delta_{\rm S}^2}{ 
\omega_{\rm c}^{2K} } \;,\quad
\epsilon_0^{2-2K} =  \frac{\Gamma^2(1-K)}{2^{2K}_{}} \frac{ \Delta_{\rm T}^2}{ 
\omega_{\rm c}^{2K}}    \; .
\]
In the scale $\epsilon_0^{}$, the selfduality of the BSG model becomes evident.
In the second form, we have anticipated from below, that
at fixed $\epsilon_0^{}$ the TSS bare coupling $\Delta_{\rm T}^{}$
is related to the BSG bare coupling $\Delta_{\rm S}^{}$ via 
 $\Delta_{\rm T}^{} = 2 \sin(\pi K)\Delta_{\rm S}^{}$ \cite{fendley95a}.

The rates of the Schmid model can be written as \cite{saleurweiss01}
\begin{equation}\label{series2}
\gamma_n^\pm = \epsilon\, \sum_{\ell=n}^\infty  x_{\rm S}^{2\ell}
\,U_{n,\ell}^\pm \;.
\end{equation}
To proceed, we map the $u$- and $v$-charges onto a single time axis. To this
end, we introduce charges $\eta_j = \pm 1$ describing 
forward/backward moves along the quasiclassical path $q+q'$ and charges 
$\xi_j = \pm 1$ representing moves along the quantum fluctuation path
$q - q'$. The cumulative charge $p_\ell = \sum_{k=1}^\ell \xi_k $
measures how far the system is off-diagonal after $\ell$ moves.
Then we have
\begin{equation}\label{useries}
U_{n,\ell}^\pm  =  \frac{1}{2^{2\ell}_{}}
\! \sum_{\{\xi_j\}'}\sum_{\{\eta_i\}'}\left[\cos\phi_\ell\,I_\ell^{+}
\pm \sin \phi_\ell\, I_\ell^{-} \right]  \; .
\end{equation}
Here $\{\cdots\}'$ denotes the constraints $\sum_j \xi_j =0 $ for the
$2\ell$ $\xi$-charges and
$\sum_i \eta_i^{} =2n$ for the $2\ell$ $\eta$-charges. 
The number of different charge sequences contributing to $U_{n,\ell}^\pm$ is 
$N_{\xi,\eta} =   \frac{(2\ell)!}{\ell! \,\ell!}
          \,       \frac{(2\ell)!}{(\ell + n)!(\ell - n)!}$.
The imaginary parts of the bath correlations in the influence functional
${\cal F}$ add up to the influence phase
$\phi_\ell^{}(\vec{\xi},\vec{\eta}) = \pi K \sum_{j=1}^{2\ell-1} 
p_j^{}\eta_j^{}  $. 

All the quantum fluctuations are in the $2\ell-1$-fold integrals
$I_{\ell}^{\pm}$, which are
independent of $\vec{\eta}$. With the notation
$  \int_0^\infty {\cal D}_{2\ell-1} (\vec{\tau}) \ldots  \equiv \int_0^\infty
{\rm d}\tau_{2\ell -1}^{}\cdots {\rm d}\tau_1^{}\ldots $, where the 
$\tau_j^{}$ are the dimensionless charge intervals (scaled with $\epsilon$), 
we get
\[
I_{\ell}^{+}(\vec{\xi}) \pm i\,I_\ell^{-}(\vec{\xi}) \equiv
\int_0^\infty  \!\!\!\!\! {\cal D}_{2\ell-1}(\vec{\tau})\,G_\ell^{({\rm c})}
(\vec{\tau},
\vec{\xi})\,e_{}^{\pm i \varphi_\ell^{}(\vec{\tau},\vec{\xi})}\; .
\]
The bias phases are combined in the phase
$ \varphi_\ell^{}(\vec{\tau},\vec{\xi}) = \sum_{j=1}^{2\ell-1}p_j^{} 
\tau_j^{} $. The interactions of the $\xi$-charges are encapsulated in the 
factor (we put $\tau_{ji}^{} = t_j - t_i = \sum_{k=i}^{j-1} \tau_k^{} \,$)
\[
G_{\ell}^{\rm (c)}(\vec{\tau},\vec{\xi}) = \exp\Big(
2K \!\!\sum_{j>i=1}^{2\ell}
\xi_j \xi_i \ln \tau_{ji}^{}\Big) \, - \, 
G_{\ell}^{\rm (subtr)}(\vec{\tau},\vec{\xi})       \; .
\]
The subtractions are to ensure that the 
$I_{\ell}^{\pm}$ represent irreducible clusters. 
Subtractions are needed whenever one or several of the 
cumulative $p$-charges are zero. For instance, putting $\ell =2$ and 
$\xi_2 = - \xi_1$, $\xi_4 = - \xi_3$ we then have
$ G_{2}^{\rm (subtr)} =  (\tau_1^{}\tau_3^{})^{-2K}_{}$.

The expression for $U_{n,\ell}^{\pm}$ can be simplified considerably
upon utilizing the following observations:
 
(i) At $T=0$, backward moves to higher wells are absent, i.e. the
aforedescribed partial backward rates $\gamma_n^{\alpha,-}$ are all zero.
Hence for any subset $\alpha$, the respective linear combination of the
$I_{\ell}^{+}$-integrals cancels that of the $I_{\ell}^{-}$-integrals,
while they add up in $\gamma_n^{\alpha,+}$. 

(ii) For every individual $\vec{\xi}$-configuration, in which all 
cumulative charges $p_j^{}$ have the same sign (some of them may be zero),
the integrals $I_{\ell}^{\pm}$ are simply related by
$I_{\ell}^{+}(\vec{\xi}) = \tan (2\pi K \ell)\, I_{\ell}^{-}(\vec{\xi})$.

Property (i) is of general nature since it is due to detailed balance.
Property (ii) is based on the scale invariant logarithmic charge 
interaction in $G_{\ell}^{\rm (c)}$ and is responsible for integrability. 
It does not hold when scale 
invariance is broken by finite cutoff or finite temperature.

Upon exploitation of properties (i) and (ii) a multitude of linear
functional relations between the various integrals 
$I_{\ell}^{\pm}(\vec{\xi})$ with different $\vec{\xi}$ 
but same $\ell$ can be derived.
The analysis yields  that all integrals $I_\ell^{\pm}(\vec{\xi})$
with one or several $p_j^{}$ equal to zero
can be expressed in terms of those which have all $p_j^{}$ nonzero.  
Making use of these relations we find that there are formidable 
cancellations among the various contributions to $U_{n,\ell}^+$.
In particular, the contributions from arrangements with any fixed nonzero 
number of  negative $\eta$-charges cancel each other exactly. 
Thus we have $U_{n,\ell}^+ = 0$ for $\ell > n$.
Hence the contributions of all paths which 
undertake backward moves between virtual intermediate states add up to zero
and only the paths with the minimal number of flips $2n$ contribute 
to the forward rate $\gamma_n^+$.
We finally get
\begin{equation}\label{res1}
U_{n,n}^{+} = \frac{(-1)^{n-1}}{n} 
\sum_{\{\xi_j\}'} I_n^{-}(\vec{\xi})
\prod_{k=1}^{2n-1} \sin(\pi K p_k^{})      \; .
\end{equation}

We now turn to the study of the incoherent dynamics in the TSS.
The subset of charge sequences in the Schmid model in which both the $u$- and 
$v$-charges alternate in sign defines a partial rate of $\gamma_1^\pm$. This
partial rate, denoted by $\tilde{\gamma}^\pm $, is the full rate describing 
incoherent relaxation in the biased TSS.
At $T=0$, the expansion in the number of transitions
(fugacity expansion) gives
\begin{equation}
\tilde{\gamma}^\pm = \sum_{\ell=1}^\infty \tilde{\gamma}_\ell^\pm \; ; \qquad 
\tilde{\gamma}_\ell^\pm = \epsilon\,x_{\rm T}^{2\ell}\, W_{\ell}^\pm \; ,
\end{equation}
where $W_{\ell}^\pm = U_{1,\ell}^\pm\,$, and where the double sum in 
(\ref{useries}) is subject to the TSS constraints.
The $\eta$-sum is easily performed, yielding
\[
W_\ell^\pm = \left(\frac{-1}{2}\right)^{\ell-1} 
\cos^{\ell}(\pi K) \!\! \sum_{ \{\xi_j\}''}
\left[ I_{\ell}^{+}  \pm \xi_1 \tan(\pi K) I_{\ell}^{-} \right] \; .
\]
Here $\{ \cdot\}''$ denotes the TSS constraint $\xi_{2k} = - \xi_{2k-1}$.

Making again extensive use of properties (i) and (ii), we find that all 
the partial backward rates $\tilde{\gamma}_n^{-}$ vanish and that $W_n^+$ 
can be expressed in terms of the
$I_n^{-}(\vec{\xi})$ integrals, in which all $p_j$ are nonzero. In the end we 
find using (\ref{res1})
\begin{equation}\label{uwrel}
W_n^+ = (-1)^{n-1} \frac{4\sin^2(n\pi K)}{[2\sin (\pi K)]^{2n}}\, 
U_{n,n}^+ \; .
\end{equation}                   
Employing relation (\ref{uwrel}), we can express the partial rate 
$\tilde{\gamma}_n^+$ of 
the TSS in terms of the rate $\gamma_n^+$ of the Schmid model.
At fixed renormalized coupling $\epsilon_0^{}$ we have
\begin{equation}\label{raterel}
\tilde{\gamma}_n^+ = (-1)^{n-1} 4\sin^2(n\pi K)  \, \gamma_n^+   \; .
\end{equation}
Upon combining the thermodynamic Bethe ansatz with the Keldysh approach, the 
rate $\gamma_n^+$ has been found as \cite{saleurweiss01}
\begin{equation}\label{rateexpl}
\gamma_n^+ = \frac{(-1)^{n-1}}{n!}\,\frac{\Gamma(\frac{3}{2})\Gamma(Kn)}{ 
\Gamma[\frac{3}{2} + (K-1)n]}\, \frac{\epsilon}{2\pi } \, v^{(2K-2)n}_{} \;. 
\end{equation}

Use of the relation (\ref{raterel}) and the expression (\ref{rateexpl})
gives the partial rate $\tilde{\gamma}_n^+$ in explicit closed form.
The perturbative expansion of the full rate $\tilde{\gamma}_{}^+$ of the TSS
is found to read
\begin{eqnarray}\label{series3}
\tilde{\gamma}^+ &=&  \frac{\epsilon}{2\sqrt{\pi}}\,\sum_{m=1}^\infty
 a_m^{}(K)\,v^{(2K-2)m}_{} \; , \\ \nonumber
a_m^{}(K) &=&
\frac{1}{m!}\, \frac{ \Gamma (Km)\,[\,1- \cos (2\pi K m)\,]}{
\Gamma[\,\frac{3}{2} +(K-1)m\,]} \,   \; .
\end{eqnarray}
Expression (\ref{series3}) is the weak-tunneling expansion of the TSS.
For rational values of $K$, the series (\ref{series3}) is  a linear
combination of hypergeometric functions.

In the regime $K<1$, the perturbative series is absolutely converging
for large enough $v$, i.e. large enough bias. For $K>1$, (\ref{series3})
is absolutely converging for small enough $v$, i.e. small enough bias.
The leading term is the Golden Rule rate
$\tilde{\gamma}_{\rm GR}^+ = [\,\pi/2\Gamma(2K)]\epsilon\, x_{\rm T}^{2}$
\cite{leggett87,weissbook}.

An integral representation for $\tilde{\gamma}^+$ is found by 
writing $\Gamma (Km)/\Gamma[\,\frac{3}{2} +(K-1)m\,]$ as a contour
integral. Within the radius of convergence of series (\ref{series3}) 
we can interchange the order of summation and integration, yielding
\[
\tilde{\gamma}^+ = {\rm Re}\,\frac{\epsilon}{2\pi i}\!\!
\int_{\cal C}^{} \! \frac{{\rm d}z}{z}
[(z-1 -z^K_{}u_2^{})^\frac{1}{2}_{}
- (z-1 -z^K_{}u_1^{})^\frac{1}{2}_{}]  \; ,
\]
where $u_1^{}=v^{2K-2}_{}$ and $u_2^{}= e^{i 2\pi K}_{} u_1^{}$.
The contour ${\cal C}$ starts at the origin, circles anti-clockwise
around the square-root branch point which  is near
$1+ u_{1/2}^{}$ for small $v^{2K-2}$, and returns to the origin. 
The integral converges for all $K$ and $v$, not just in the 
regime where the series (\ref{series3}) does. Therefore it must also yield 
the asymptotic expansions for small (large) enough 
$v$ when $K < 1$ ($K > 1$).

The asymptotic expansion for $K < 1$ can be found by 
changing variable to $y= (u_{1/2}^{}z)^{1/(1-K)}_{}$ and 
expanding the integrand for small $v$. The strong-tunneling series is
\begin{equation}\label{asymp1}
\tilde{\gamma}^+ = 
\frac{\epsilon}{2\sqrt{\pi} } \sum_{n=0}^\infty\, b_n^{}(K) 
\,v^{2n-1}_{} \; .
\end{equation}
For $ \frac{1}{3} < K < 1 $, the branch cut of the $u_2^{}$-term is not 
encircled by the contour. In this regime we obtain
\begin{equation}\label{as1} 
b_n^{}(K) = d_n^{}(K) \equiv
\frac{1}{ n!}\,\frac{\Gamma[(\frac{1}{2}-n)
\frac{K}{1-K}]}{(\frac{1}{2}-n)\Gamma[(\frac{1}{2}-n)\frac{1}{1-K}]} \;.
\end{equation}
For $K \le \frac{1}{3}$, both 
the $u_1^{}$- and $u_2^{}$-term contribute, giving
\begin{equation}\label{as2}
b_n^{}(K) = 2\sin^2[\textstyle{\frac{\pi K}{1-K}}(\frac{1}{2}-n)]
\,d_n^{}(K) \; .
\end{equation}

In the limit $K\ll 1$, the perturbative series (\ref{series3}) 
as well as the asymptotic series (\ref{asymp1}) can be summed to the
known result $\tilde{\gamma}^+_{} = \pi K \Delta_{\rm T}^{2}\Big/
\sqrt{\Delta_{\rm T}^2 +\epsilon^2_{}} \,$ \cite{weissbook}. 

The asymptotic series (\ref{asymp1}) is an expansion in powers of 
$\epsilon^2_{}/\epsilon_0^2$. The leading term is independent of $\epsilon$
and represents the forward rate for the symmetric TSS,
\begin{equation}\label{gamma0}
\tilde{\gamma}_0^+ = (2\sqrt{\pi}\, )^{-1}_{} \,b_0^{}(K)\,\epsilon_0^{}\; , 
\end{equation}
where $b_0^{}(K)$ is given in (\ref{as1}) and (\ref{as2}), respectively.

\begin{figure}[t]
\includegraphics[width=73mm]{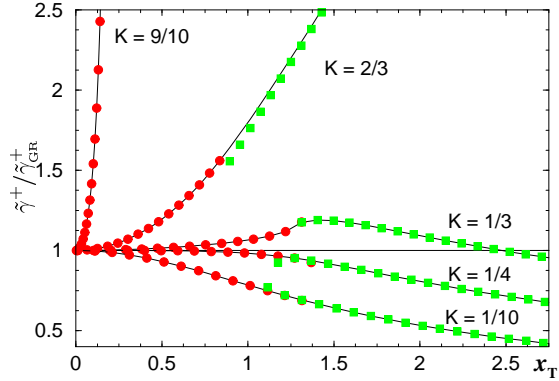}
\vspace{-3mm}
\caption{The scaled rate
$\tilde{\gamma}^+_{}/\tilde{\gamma}_{\rm GR}^+$ is plotted versus
$x_{\rm T}^{}$ for various $K<1$.
The circles represent the weak-tunneling
series and the squares the strong-tunneling expansion. The full curve is
the hypergeometric function expression
\label{figone}}
\end{figure}

In Fig.~\ref{figone}, the scaled rate 
$k^+_{} \equiv \tilde{\gamma}^+_{}/ \tilde{\gamma}^+_{\rm GR}$
is plotted for different $K < 1$. The horizontal line is the 
$K=\frac{1}{2}$ result. For $K < \frac{1}{4}$, the various tunneling
contributions interfere always destructively, which leads to a reduction of 
the rate. For $\frac{1}{2} < K < 1$,  the tunneling
contributions interfere constructively for all $x_{\rm T}^{}$. 
In the range $\frac{1}{4} < K < \frac{1}{2}$, the rate goes through a maximum.
This reflects constructive interference  at small and 
intermediate $x_{\rm T}^{}$, and destructive interference at large 
$x_{\rm T}^{}$.

The nonperturbative result (\ref{gamma0}) for the relaxation rate
may be compared with the decoherence rate $\gamma_{\rm dec}^{}$ describing
damping of the coherent oscillations in the TSS \cite{weissbook}. This rate 
has been calculated for the symmetric TSS in Ref.~\cite{lesage97} within the
framework of integrable QFT. In our notation, the result is
$\gamma_{\rm dec}^{} = \tilde{\gamma}_0^+/2$ for $K\le \frac{1}{3}$ and
$\gamma_{\rm dec}^{}=\sin^2[\frac{\pi K}{2(1-K)}]\tilde{\gamma}_0^+$
for $\frac{1}{3} < K < \frac{1}{2}$. It is known that the same relations
hold for a biased TSS in the regimes $K\ll 1$ and $K$ close to $\frac{1}{2}$
\cite{weissbook}. Upon combining these relations with 
Eqs.~(\ref{asymp1}) - (\ref{as2}) it is natural to conjecture that the
decoherence rate of a biased TSS is
\begin{equation}\label{deco}
\gamma_{\rm dec}^{} = \frac{\epsilon}{2\sqrt{\pi}}\sum_{n=0}^\infty
{\textstyle \sin^2[\frac{\pi K}{1-K}(\frac{1}{2}-n)]}
\,d_n^{}(K)\, v_{}^{2n-1}\;,
\end{equation}
in agreement with all the known results in special cases.
This expression gives the lower ($T=0$) bound for quantum
decoherence in the entire regime  $0 <K \le \frac{1}{2}$.
\begin{figure}[t]
\includegraphics[width=72mm]{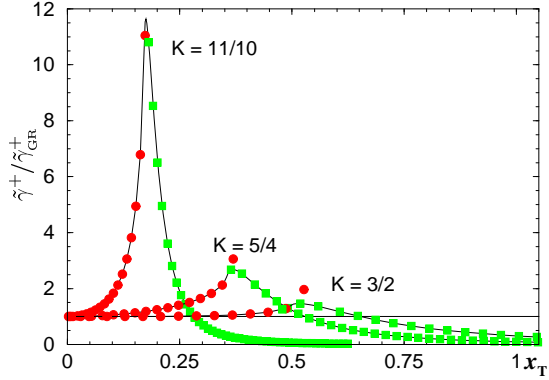}
\vspace{-3mm}
\caption{The same scaled rate, but now for various $K > 1$. \label{figtwo}}
\end{figure}

Finally consider the damping regime $K>1$ for large $v$ (large bias).
By defining the variable $t = e^{-i\pi}_{}u_{1/2}^{}z^K_{}$, we may 
expand the resulting integrand in powers of $u^{-1/K}_{1/2}$.
Introducing the sections $K= p + \kappa$ with $p=1,2,\cdots$ and 
$0 \le \kappa < 1$, we obtain the asymptotic series as
\begin{eqnarray}\label{asymp2}
\tilde{\gamma}^+ &=& \frac{\epsilon}{2\sqrt{\pi} }\, 
\sum_{m=1}^\infty c_m^{}(K)\, v^{(2/K - 2)m}_{}  \; , \\[2mm]  \nonumber
c_m^{}(K) &=& 
\frac{(-1)^{m}_{}}{m!}\,\frac{2\, \Gamma(\frac{m}{K})
\sin[\,\frac{1+p}{K} m\pi] 
\sin(\frac{p}{K}m\pi)}{ K\, \Gamma[\,\frac{3}{2} 
+  (\frac{1}{K} -1 )m\, ]} \; .
\end{eqnarray}
This asymptotic series (\ref{asymp2}) bears a strong resemblance with that of 
the self-dual Schmid model \cite{fisher85,fendley95b,fendley98}. 
The powers follow from those 
of the perturbative expansion (\ref{series3}) by the substitution $K\to 1/K$, 
as also do major parts of the coefficient $c_m^{}(K)$. 
However, it differs from 
self-duality by the alternating sign and by the sine-factors.

Fig.~\ref{figtwo} gives plots of the scaled rate $k^+_{}$
for $K>1$. The
rate goes through a maximum which is shifted to higher $x_{\rm T}^{}$ with
increasing $K$. At large enough $x_{\rm T}^{}$, the strong-tunneling
contributions interfere always destructively. 



\begin{thebibliography}{199}

\bibitem{weissbook}
U. Weiss, {\em Quantum Dissipative Systems} (World Scientific,
Singapore, 2nd edition, 1999).

\bibitem{fendley95a}
P. Fendley and H. Saleur, Phys. Rev. Lett. {\bf 75}, 4492 (1995). 

\bibitem{leggett87}
A.J.~Leggett, S.~Chakravarty, A.T.~Dorsey, M.P.A.~Fisher, A.~Garg, and
W.~Zwerger, Rev. Mod. Phys. {\bf 59}, 1 (1987).

\bibitem{schmid83}
A.~Schmid, Phys. Rev. Lett. {\bf 51}, 1506 (1983).

\bibitem{saleurweiss01} 
H. Saleur and U. Weiss, Phys. Rev. B {\bf 63} 201302(R) (2001).

\bibitem{fendley95b}
P. Fendley, A. Ludwig, and H. Saleur, Phys. Rev. B {\bf 52}, 8934 (1995).


\bibitem{lesage97}
F. Lesage and H. Saleur, Phys. Rev. Lett. {\bf 80}, 4370 (1998). 

\bibitem{fisher85}
M.P.A. Fisher and W. Zwerger, Phys. Rev. B {\bf 32}, 6190 (1985).

\bibitem{fendley98}
P. Fendley and H. Saleur, Phys. Rev. Lett. {\bf 81}, 2518 (1998).
\end{thebibliography}
\end{document}